\documentclass[12pt,floats]{article}
\usepackage{epsfig}
\begin{document}

\title{\bf How to model BEC numerically?}

\author{Oleg V. Utyuzh$^{1}$, Grzegorz Wilk$^{1}$ and 
Zbigniew W\l odarczyk$^{2}$  \\
$^{1}${\it The Andrzej So\l tan Institute for Nuclear Studies,}\\
      {\it Ho\.za 69, 00681 Warsaw, Poland,} \\
      {\it e-mail: utyuzh@fuw.edu.pl }\\
$^{2}${\it Institute of Physics, \'Swi\c{e}tokrzyska Academy,} \\
      {\it \'Swi\c{e}tokrzyska 15, 25-405 Kielce, Poland,} \\
      {\it e-mail: wlod@pu.kielce.pl }
      }
\date{\today}
\maketitle

\begin{abstract}
The new method of numerical modelling of Bose-Einstein correlations
observed in all kinds of multiparticle production processes is
proposed. \\

\end{abstract}

\newpage

The quantum mechanical in its origin Bose-Einstein correlations (BEC)
between identical bosons are our basic source of the knowledge on
the space-time characteristics of the hadronizing objects formed  
in high-energy collisions \cite{Weiner}. Despite the long history of
the BEC study the question of their numerical modeling of some
Monte-Carlo (MC) event generators remains still open. This is because
MC event generators are ({\it by construction}) probabilistic in
nature. One is therefore usually using some method to enhancing some
final configurations of momenta of observed secondaries to enhance
the observed amout of pairs of identical bosons with small mementa
differences \cite{Weiner}. Here we would like to advocate different
approach using observation made \cite{KZ} that symetrization of the
amplitudes of system of identical bosons leads to geometrical
distribution of the like-particles in the phase-space cell they
occupy (cf. Table I). There are two possible ways to implement this
idea numerically.
\begin{table}[ht]
\caption{Boltzmann vs BEC}
\begin{center}
\begin{tabular}{|ccc|} \hline 
 {\bf Boltzmann} &  & {\bf BEC }\\ 
\hline
\hline
{$\Psi_N = \prod_i \psi_i(x_i)$ } & {\small $ SYMETRIZATION $}  &
{$\Psi_N = \frac{1}{N!} \sum_{{\mathcal{P}}\{i,j\}} \prod_i  \psi_i(x_j)$ }\\ 
\hline
\hline
 {\it POISSONIAN} & & {\it GEOMETRICAL}\\
\hline
\hline 
{\quad  $P_{Boltzmann}(N) = \frac{\nu^N}{N!} e^{-\nu}$ }&
{$ \times N! $}  & {$P_{BEC}(N) = (1-\nu)\nu^N$} \\ 
\hline
\end{tabular}
\end{center}
\end{table} 
\noindent
The first one (see \cite{BECPL,BECAP} for details) is a kind of
"afterburner", which can be used with any MC event generator
providing us with distribution of particles. The idea is to use the
phase-space distribution as given by this MC but to allocate anew
charges to all particles (conserving the number of $+/-$ and neutrals
as given by MC) in such a way as to effectively put some particles to
the same cell (in which they are distributed according to geometrical
distribution). Here we would like to go step further and present a MC
event generator that produces particles according to the
Bose-Einstein statistics. This can be done
in the following way (cf.  Fig.1). Particles (pions in our case) are
selected from the pool of energy $W$ according to a thermal
distribution $f(E) = \exp(-E/T)$ and allocated to a phase-space cell
given by its $4-$momentum $p$ (we call it elementary emitting cell
(EEC)). New particles of the same charge and energy are then added to
this cell according to $P=P_0 \exp(-E/T)$ weights until the first
failure\footnote{It should be stressed that such choice results
immediately in Bose-Einstein character of the cell occupancy,
$\langle n\rangle = P/(1-P) = 1/[\exp(E/T)/P_0-1] = 1/[\exp((E+\mu)/T -1]$
where $\mu = \ln P_0$.}. The process  continues with formation of new
EEC's until the initial energy $W$ is used up. At the end one
corrects for energy-momentum and charge conservation. In this way:
$(i)$ distribution of EEC's is poissonian and $(ii)$ distribution of
particles (of the same charge) in a cell is geometrical. This results
in a Negative Binominal (NB) form of the over all multiplicity
distribution.   

\begin{center}
\begin{tabular}{|cc|}
\hline
\begin{minipage}{5cm}
\begin{center}
\includegraphics[height=3cm,width=5cm]{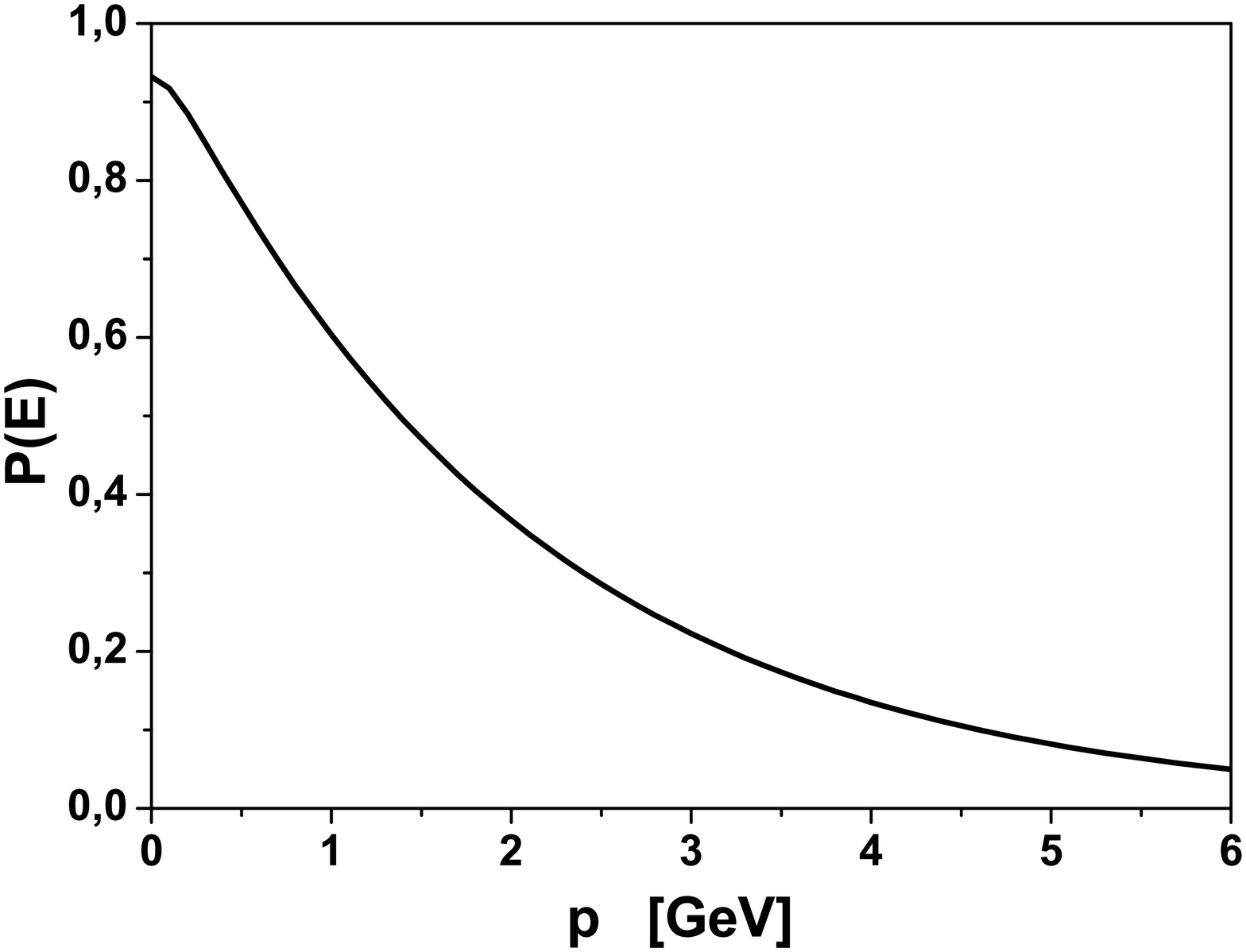} 
\end{center}
\end{minipage} &
\begin{minipage}{5cm}
\begin{center}
\includegraphics[height=3cm,width=5cm]{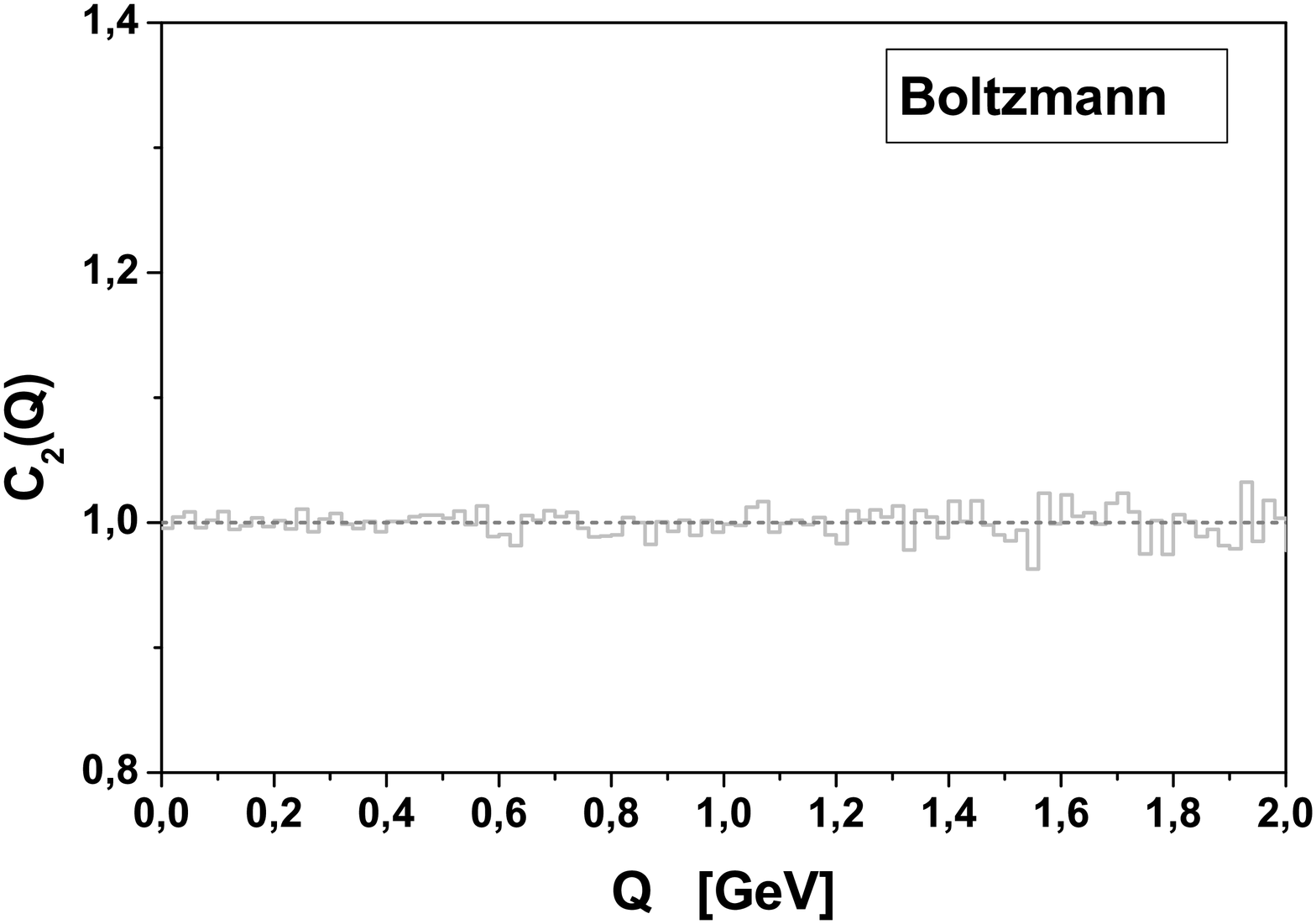}
\end{center}
\end{minipage} \\
\begin{minipage}{5cm}
\begin{center}
\includegraphics[height=3cm,width=5cm]{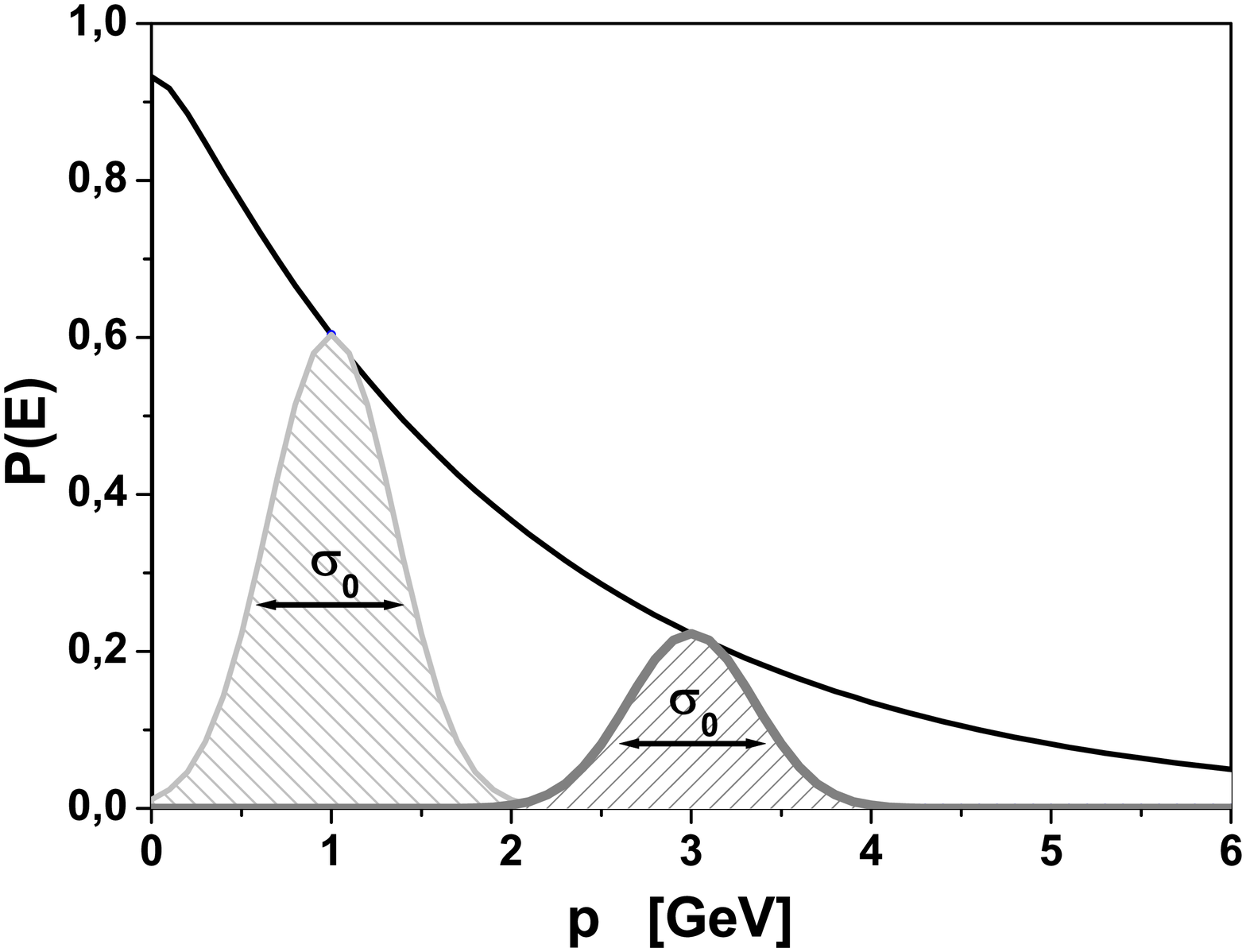}
\end{center}
\end{minipage} &
\begin{minipage}{5cm}
\begin{center}
\includegraphics[height=3cm,width=5cm]{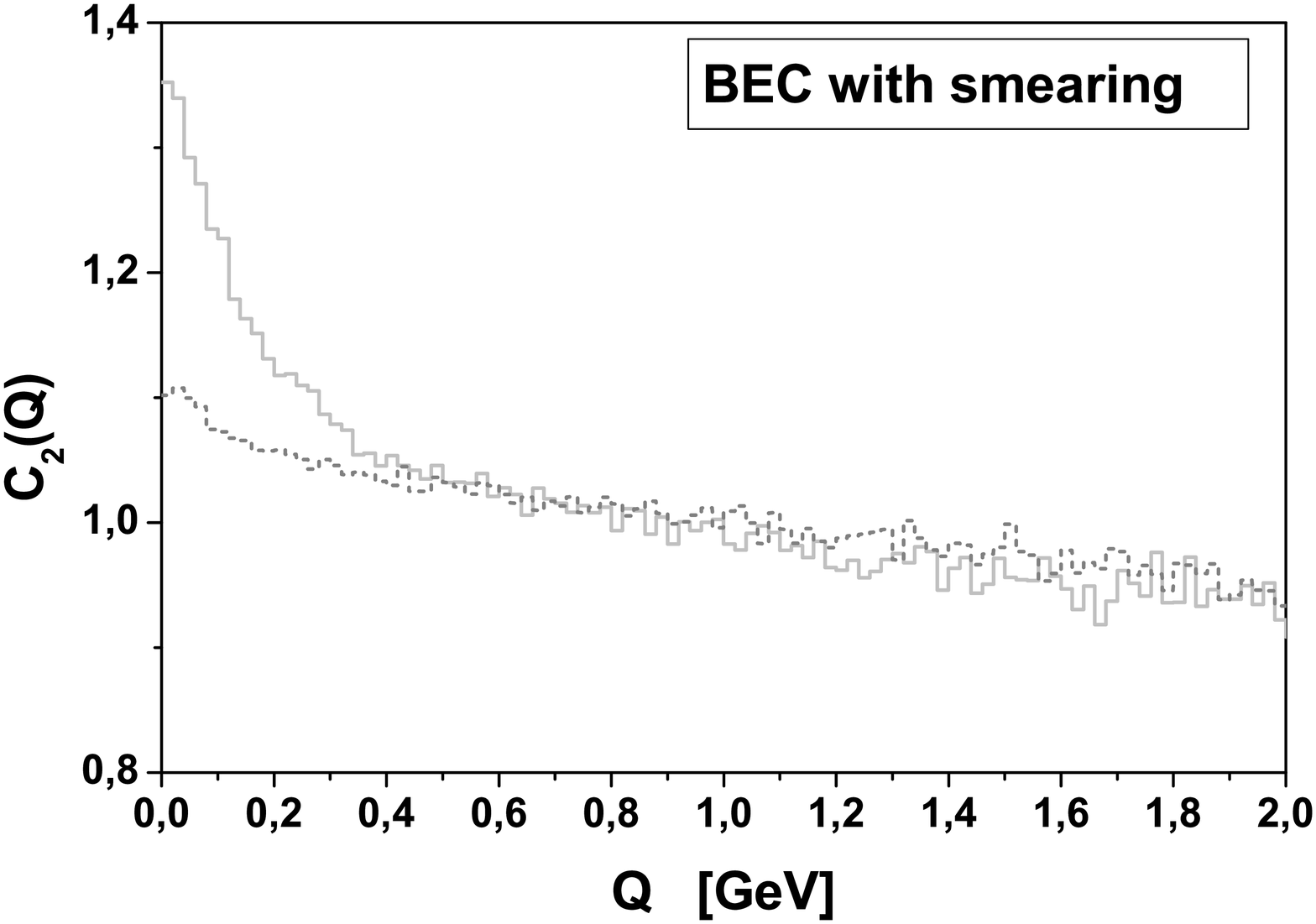}
\end{center}
\end{minipage} \\
\hline
\end{tabular} \\ 
\vspace{0.5cm}{\scriptsize Fig.1 The particles selected according to
thermal distribution (upper-left) show no BEC (upper-right).
Correlations occur only (lower-right) when one allows more particles
at the same cell (lower-left). The shape of the correlation function
$C_2(Q)$ is highly correlated with the spread of the cells in the
momentum space.} 
\end{center}
It turns out that in order to get proper observed structure of
correlation function $C_2(Q)$ (and to be able to describe the
experimental data, cf. Fig. 2) one has to allow for some smearing of
momenta of like particles allocated to a given EEC (here given, for
example, by a gaussian form with width $\sigma$ defining the size of
average EEC).

\begin{center}
\begin{tabular}{|cc|}
\hline
\begin{minipage}{5cm}
\begin{center}
\includegraphics[height=5cm,width=5cm]{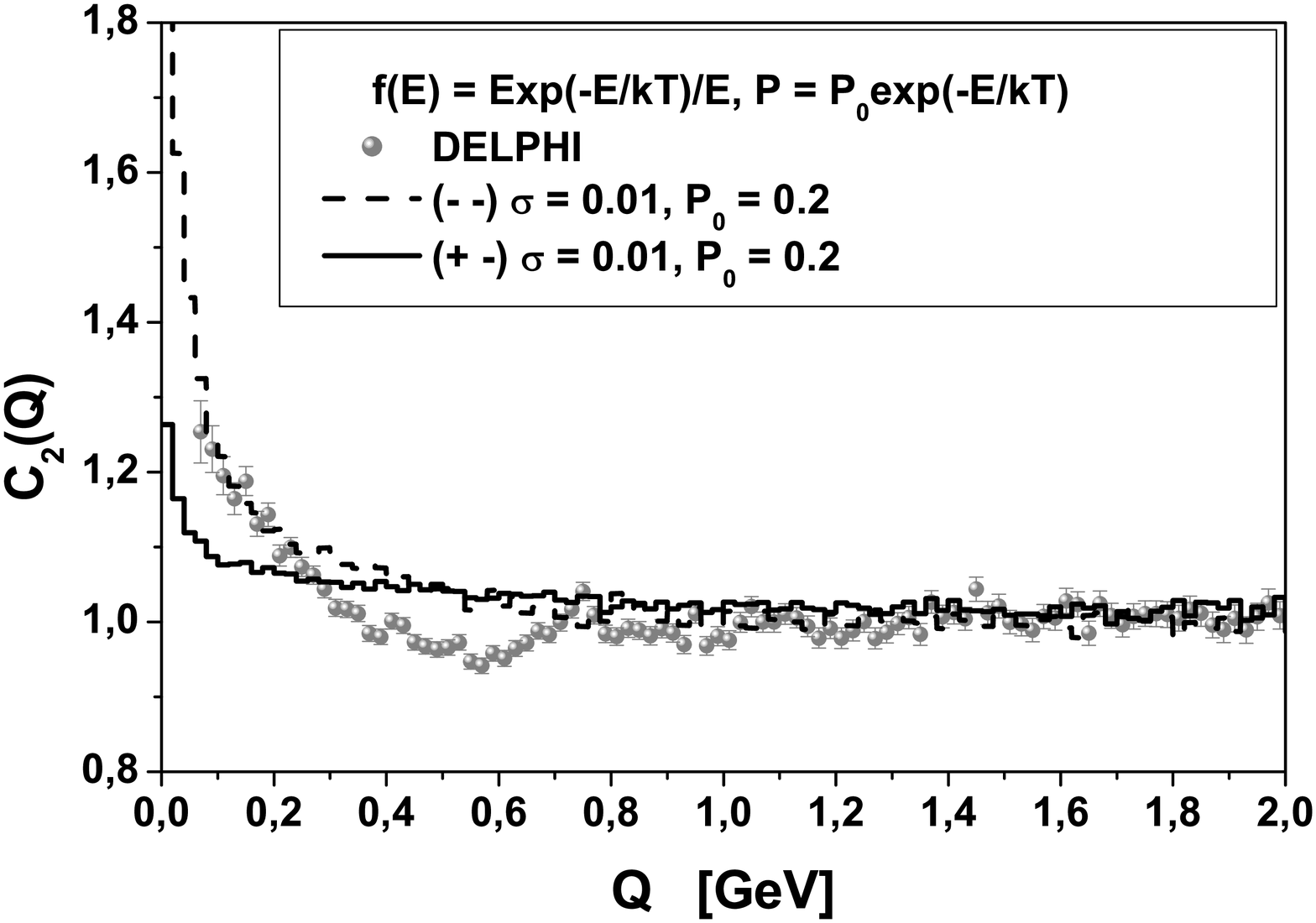} 
\end{center}
\end{minipage} &
\begin{minipage}{5cm}
\begin{center}
\includegraphics[height=5cm,width=5cm]{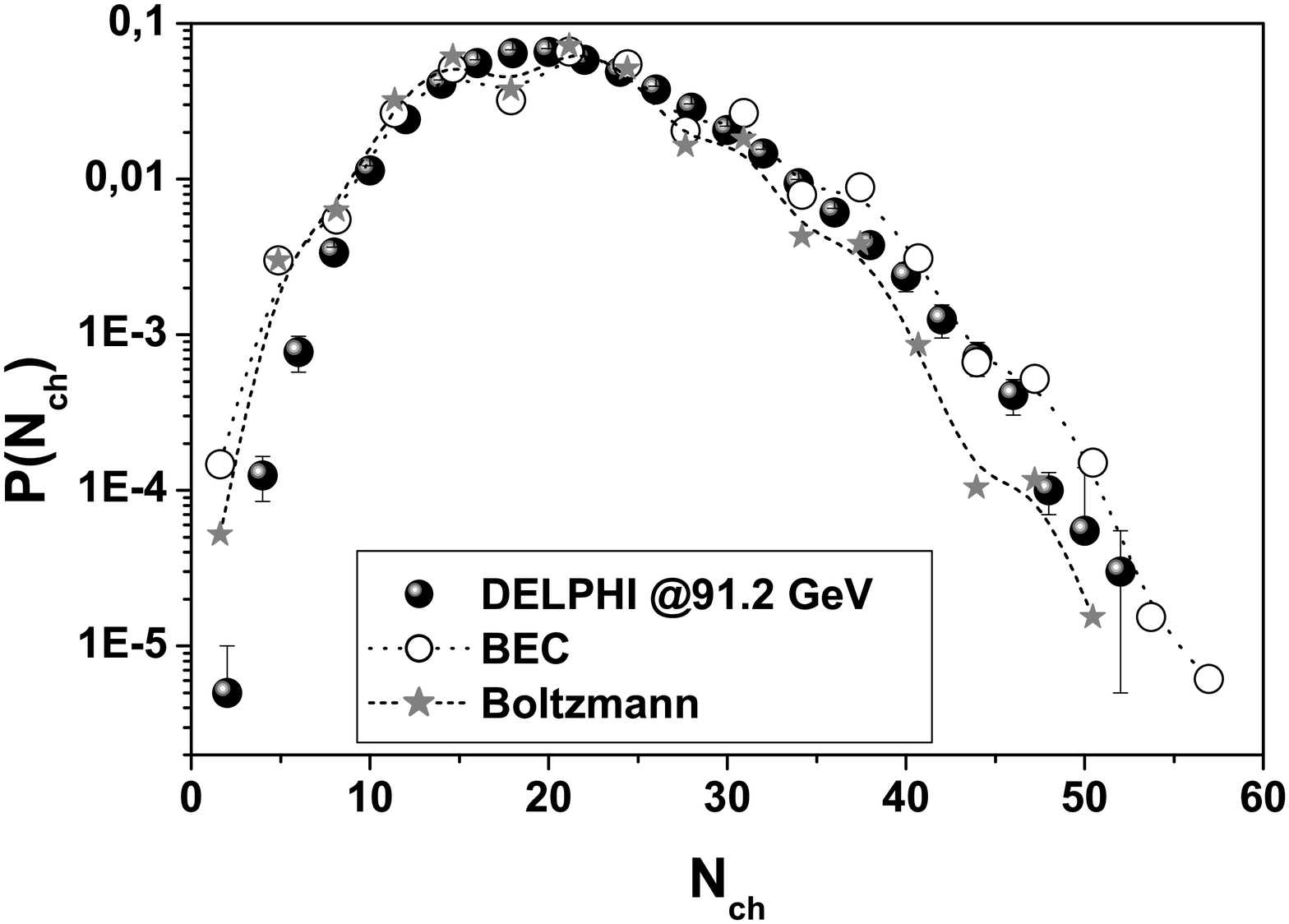}
\end{center}
\end{minipage} \\
\begin{minipage}{5cm}
\begin{center}
\includegraphics[height=4cm,width=5cm]{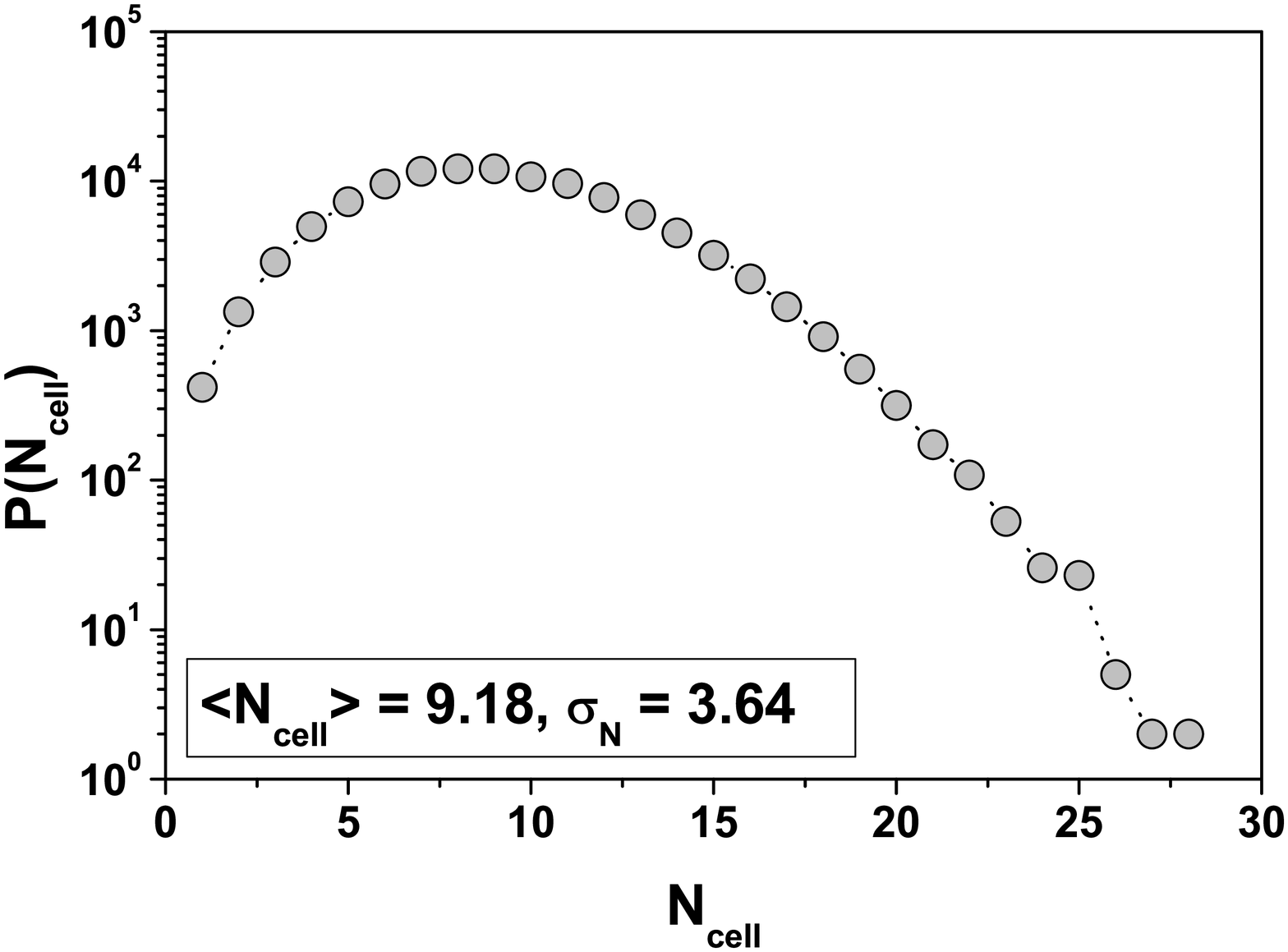}
\end{center}
\end{minipage} &
\begin{minipage}{5cm}
\begin{center}
\includegraphics[height=4cm,width=5cm]{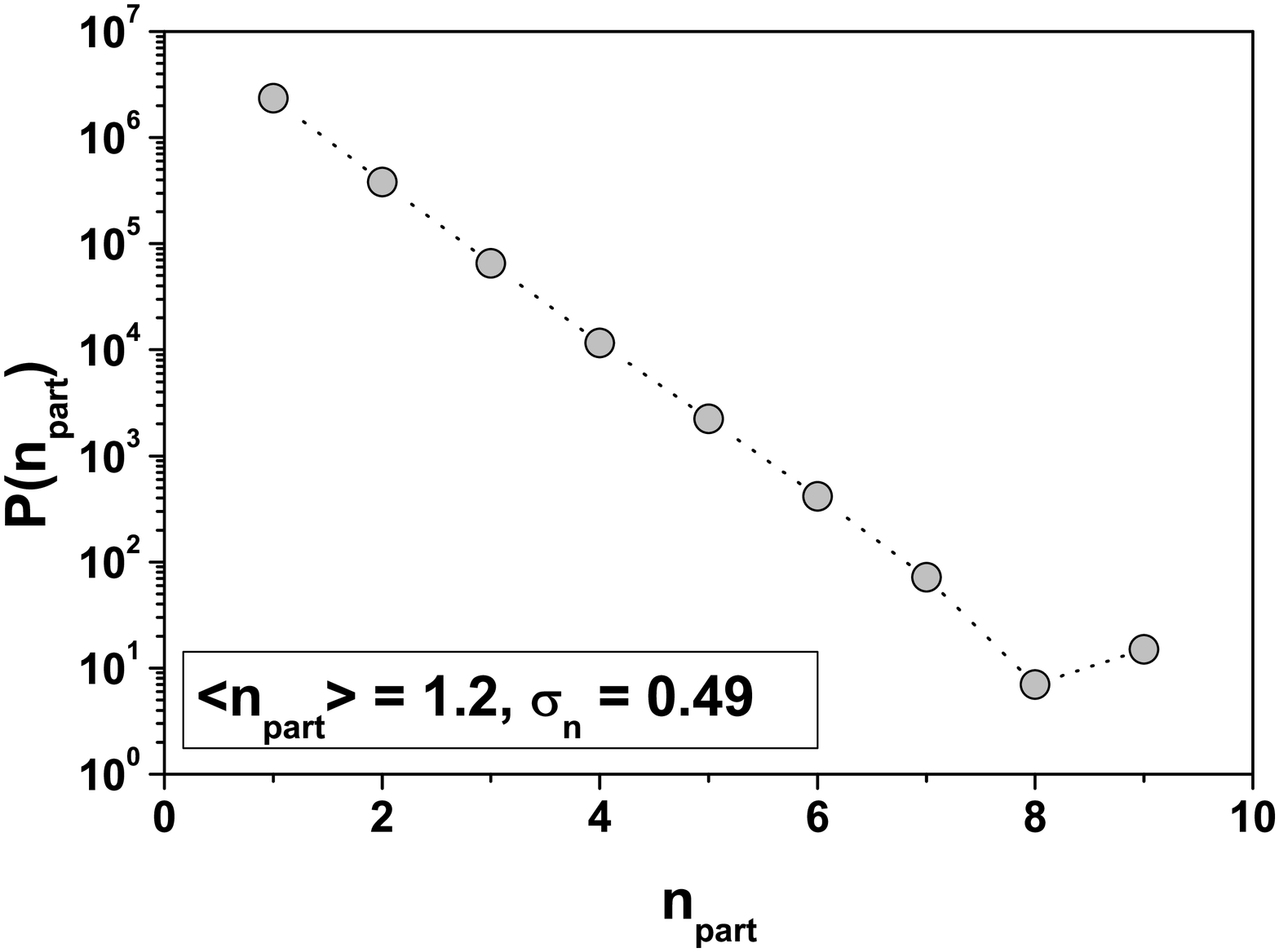}
\end{center}
\end{minipage} \\
\hline
\end{tabular} \\ 
\vspace{0.5cm}{\scriptsize Fig.2 Example of calculations showing
that we can fit $e^+e^-$ data \cite{Data} with some choice of parameters
(upper-left) and at the same time corresponding multiplicity
distribution (upper-right). Lower panels show corresponding
distributions of the number of EEC's (which has poissonian
distribution shape - left panel) and distribution of number of
particles in EEC (which has geometrical distribution shape - right
panel).}  
\end{center}
In Fig. 2 we show some example of results for BEC obtained this way
(including a comparison with experimental data for BEC observed in
$e^+e^-$ annihilations \cite{Data}) together with the corresponding
results for multiplicity distributions, distributions of the
number of formed EEC's and distributions of particles in them.
Although the fit to data is quite reasonably one should treat results
presented here only as a necessary first proof that such program is
in principle possible to be implemented and that it is flexible to
address (so far only some) experimental data\footnote{Our approach is
still essentially one-dimensional with transverse momenta entering
only via their mean value, $\langle p_T\rangle$, all produced
particles are assumed to be $\pi^{(+,-,0)}$ only (no resonances are
considered either), no corrections for any form of final state
interaction is attempted. All these factors has to be included before
one can use this method to serious description of experimental data
for all kind of reactions. We plan to pursue such research in the
future.}.    

We would like to close with the following remarks. The approach
presented here can be traced to first attempts of describe effect of
BEC, only later came space-time descriptions which dominate at
present and which use symmetrization of the corresponding wave
functions with the space-time integration over some assumed source
function (i.e., function describing space-time distribution of points
in which finally observed secondaries are produced\footnote{By this
we understand points after which they no more interact among
themselves.}) \cite{Weiner,KZ}. The shape of this source function
consists then the main object of investigation. From our approach it
is obvious that this shape emerges from the necessary spread in the
momenta (cf. Fig. 1) and therefore is not so much given by the
dimension of the source as by the length at which particles could
still be considered as belonging to a given cell (cf. \cite{Weiner}
where such distinction has been also introduced but on different
grounds).


\begin{thebibliography}{99}

\bibitem{Data} Abreu P et al. (DELPHI Collab.) (1992) Bose-Einstein
               correlations in the hadronic decays of the $Z^0$.
               Phys Lett B 286:201-210.

\bibitem{BECPL} Utyuzh O, Wilk G, W\l odarczyk Z (2001) Numerical
                modelling of Bose-Einstein correlations. Phys Lett B
                522:273-279.

\bibitem{BECAP} Utyuzh O, Wilk G, W\l odarczyk Z (2002) Bose-Einstein
                correlation as reflection of correlations of
                fluctuations. Acta Phys Pol B 33:2681-2693.

\bibitem{Weiner} Weiner R (2000) Boson interferometry in high energy
                 physics. Phys Rep 327:249-346.

\bibitem{KZ} Zalewski K (1999) Bose-Einstein correlations in
             multiparticle production. Nucl Phys. Proc Suppl 74:65-67.

\end{thebibliography}
\end{document}